\documentclass[aps,prl,twocolumn,groupedaddress,showpacs]{revtex4}
\usepackage[utf8]{inputenc}
\usepackage[english]{babel}
\usepackage{amsmath}
\usepackage{amssymb}

\renewcommand{\div}{\mathrm{div}\,}
\DeclareMathOperator{\rot}{rot}
\newcommand{\vek}[1]{\mathbf{#1}}

\newcommand{\upd}{\mathrm{d}}	
\newcommand{\ii}{\mathrm{i}}	
\newcommand{\e}{\mathrm{e}}		
\newcommand{\tens}[1]{\hat{#1}}	
\renewcommand{\Im}{\mathrm{Im}\,}

\newcommand{\Lh}{\frac{L}{2}}

\newcommand{\E}{\vek E}
\newcommand{\B}{\vek B}
\renewcommand{\H}{\vek H}
\newcommand{\D}{\vek D}
\newcommand{\M}{\vek M}
\renewcommand{\P}{\vek P}
\newcommand{\X}{\vek X}
\newcommand{\Y}{\vek Y}
\newcommand{\jind}{\vek{j}_{\mathrm{ind}}}
\newcommand{\jext}{\vek{j}_{\mathrm{ext}}}
\newcommand{\elec}{_{\mathrm{elec}}}
\newcommand{\magn}{_{\mathrm{magn}}}

\newcommand{\ie}{\textsl{i.e.}}
\newcommand{\eg}{\textsl{e.g.}}

\begin{document}

\title{Poynting's theorem and energy conservation\\in the
propagation of light in bounded media}

\author{Felix Richter}
\email{felix.richter2@uni-rostock.de}
\author{Matthias Florian}
\author{Klaus Henneberger}

\affiliation{Universit\"at Rostock, Institut f\"ur Physik, 18051 Rostock, Germany}

\date{December 17, 2007}

\begin{abstract}
Starting from the Maxwell-Lorentz equations, Poynting's theorem is reconsidered.
The energy flux vector is introduced as $\mathbf{S}_{e}=(\mathbf{E}\times\mathbf{B})/\mu_0$
instead of $\mathbf{E}\times\mathbf{H}$, because only by this choice the energy dissipation
can be related to the balance of the kinetic energy of the matter subsystem. 
Conservation of the total energy as the sum of kinetic and electromagnetic energy follows.
In our discussion, media and their microscopic nature are represented exactly
by their susceptibility functions, which do not necessarily have to be known. 
On this footing, it can be shown that energy conservation in the propagation 
of light through bounded media is ensured by Maxwell's boundary conditions alone, 
even for some frequently used approximations. This is demonstrated for
approaches using \emph{additional boundary conditions} and the \emph{dielectric
approximation} in detail, the latter of which suspected to violate energy
conservation for decades.
\end{abstract}


\maketitle

\section{Introduction}

The effects of a medium on radiation (\eg,\ in an experiment on the 
transmission and reflection of light in a dielectric medium) are determined exactly 
by the susceptibility function $\tens\chi(\vek{r},\vek{r'},t,t',\vek{E})$, 
which enters Maxwell's equations through the material equations 
(also known as constitutive relations).
Due to its complexity, it is only in rare cases that $\tens\chi$ can be expressed
analytically or valuable relations can be drawn from it. For a long time, the
more complex properties of $\tens\chi$ have not been of much interest, and textbooks
still introduce the susceptibility as a constant scalar or as a constant tensor,
to account at least for anisotropy (\eg, \cite{Born2003,Landau}). 
Non-linearity (\ie, dependence on $\vek{E}$) is an own interesting field
of research and technology but out of scope here and will be neglected. 

But strictly spoken, the presence of a medium boundary alone enforces 
a non-local and spatially inhomogeneous $\tens\chi$, if spatial dependence
in the dispersion relation (\emph{spatial dispersion}) is not explicitly neglected.
In crystal and semiconductor optics, 
some approaches to this problem resort to using the susceptibility of an 
infinite medium for the description of a bounded real one. 
Then, multiple polariton modes are predicted to propagate concurrently in the medium, 
and the precondition of a continuous transition of the electromagnetic fields at the medium boundaries 
(\emph{Maxwell's boundary conditions}, MBC's) is no longer sufficient to
determine the amplitudes and phases of these modes. 
As a solution, different phenomenological \emph{additional boundary conditions} (ABC's)
were proposed \cite{pekar-abc,ting}, the first of which in 1957. 
In 1973, a related approach was published \cite{Maradudin1973}, 
suggesting to limit the bulk susceptibility to the bounded medium by step functions 
(\emph{dielectric approximation}, DA). 
These methods were applied in a vast number of publications in the past decades,
and they still are used today and sucessfully explain experimental data
\cite{Seemann2006}. 

However, it is known that some of these approximations can conflict with the fundamental
law of energy conservation \cite{BishopMaradudin76,Venger2004}. 
Thus, scientists were prompted to take greatest care in their argumentation or 
even to reject these approximations despite of their successes (see, \eg,
\cite{Venger2004,MuljarovZimmermann}). 

We will show that, in the case of the DA and other important approximations,
energy conservation is not a question of the approximation but rather is ensured by the
MBC's alone. The DA is addressed in detail: The compatibility of the approximation with 
energy conservation is proven, and problems in the respective work are pointed
out. For all this, we use an approach that one of us recently developed
\cite{planckslaw}, which provides analytically exact relations between radiation
and matter in the elusive case of spatial dispersion with the aid of Keldysh's
photon Green's functions. 

Throughout the discussion of bounded media, we regard arbitrary steadily
excited media. They are represented by their susceptibility functions, 
which do not necessarily have to be known. Besides the latter, 
the cornerstone of our discussion is \emph{Poynting's theorem} (PT), 
the equivalent of the energy conservation law in electrodynamics. 
It balances the time derivative of electromagnetic and 
mechanical energy $U_{e,m}$ and sources of respective energy flux 
$\vek{S}_{e,m}$:
\begin{align}
\label{pt-general}
\frac{\partial}{\partial t}\left( U_e + U_m\right) + \div \left( \vek{S}_e +
\vek{S}_m\right) = 0
\end{align}
In Poynting's original work \cite{Poynting1884} and many established textbooks
\cite{Landau,Born2003,Jackson}, the theorem is given---without regarding
spatial dispersion---as
\begin{align}
\label{eq-vekpoynting-e-old}
\E\dot\D + \H\dot\B + \div \vek S_e = -\jext\E, \quad
\vek S_e = \div (\E \times \H).
\end{align}
An expression for the electromagnetic energy density $U_e$ can only be
given with further assumptions\footnote{%
Landau/Lifshitz \cite{Landau} only give the differential $\upd U_e = \E\upd \D +
\H\upd \B$; Born and Wolf \cite{Born2003} give 
$U_e = \left(\varepsilon_0\tens\varepsilon E^2 + 
\mu_0\tens\mu H^2 \right)/2$, valid for material equations \eqref{bh-hist-symmetry}; 
Jackson \cite{Jackson} gives 
$U_e = \left(\E\D + \H\B \right)/2$ for the multipole approach.}. This is why
eq.~\eqref{eq-vekpoynting-e-old} cannot be generally interpreted as an energy
continuity equation and, consequently, $\div (\E \times \H)$ not as an energy
flux vector. 

In order to lay solid grounds for our considerations, we will at first give a
thorough derivation of the theorem \eqref{pt-general} in conjunction with spatial dispersion and
mechanical energy, giving hints on why several articles, such as
\cite{NelsonGeneralizingPoynting} or those referenced therein, 
doubt the validity and form of eq.~\eqref{eq-vekpoynting-e-old} in more complex
situations.

A note on terminology: Historically, the term \emph{Poynting vector} denoted
the electromagnetic energy flux $\vek{S}_e$ only. With regard to 
eq.~\eqref{pt-general}, we here need to distinguish clearly between 
the \emph{electromagnetic} and the \emph{mechanical Poynting vector}.

\section{A clean-room derivation of Poynting's theorem}

There are many doubts expressed on under which conditions (static magnetic
fields, stationary situations, dispersive media) PT applies and on how
Poynting's energy flux vector has to be interpreted 
(compare \cite{aboutPT,Nelson2,NelsonGeneralizingPoynting,JiangLiu}). 
Here, we do not see any reason for these. Instead, we will give a brief
derivation of PT, while carefully accounting for any approximations made or
conditions implied.
This derivation is done in classical physics but can easily be extended to
quantum physics by applying the well-known recipes 
(\ie, operators, symmetrization, commutation relations).

Surprisingly, we do not need any assumptions about the matter. 
Avoiding constitutive relations and derived fields, we rather rely on 
the \emph{microscopic} Maxwell-Lorentz equations \cite{Cohen-Tannoudji1989}, 
which are the most fundamental and universal: 
\begin{subequations}
\label{maxwell-lorentz}
\begin{align}
	\label{maxwell-micro}
	\rot\E = -\dot\B, \quad \rot\B = \mu_0\vek{j} + \frac{1}{c^2}\dot\E, \\
	\label{particle-current}
	\vek{j}(\vek r, t) = \sum_{\alpha} e_{\alpha}
	\dot{\vek{r}}_{\alpha}\delta(\vek{r}-\vek{r}_{\alpha}(t)),
\end{align}
\end{subequations}
with $\vek j$ as a current of charged particles $\alpha$. 

With the vector identity $\div (\X\times\Y)=\Y\rot\X-\X\rot \Y$,
we immediately have 
\begin{align}
	\label{pt-em}
	\varepsilon_0 \E\dot\E + \frac{1}{\mu_0}\B\dot\B +
	\frac{1}{\mu_0}\div(\E\times\B) = -\vek{j}\E
\end{align}
and may identify the electromagnetic field energy density and the
electromagnetic Poynting vector as 
\begin{align}
\label{eq-energydensity-e}
U_{e}&=\frac{1}{2}(\varepsilon_0\E^2 + \frac{1}{\mu_0}\B^2),\\
\label{eq-vekpoynting-e}
\vek{S}_{e}&=\frac{1}{\mu_0}(\E\times\B).
\end{align}
Note that in what follows, we will use these new definitions of the two
quantities instead of the more conventinal given in eq.~\eqref{eq-vekpoynting-e-old},
if not otherwise stated.

Now, a microscopic mechanical expression for $\vek{j}\E$ shall be
developed. We use the density of mechanical (kinetic) energy of the particles
$\alpha$, $U_m$, as well as their kinetic energy flux, $\vek{S}_m$, according
to 
\begin{align}
U_m(\vek{r},t) &= \sum_{\alpha} \frac{m_{\alpha}}{2} \dot{r}^2_{\alpha}
\delta(\vek{r}-\vek{r}_{\alpha}(t)),\\ 
\vek{S}_m (\vek{r},t) &= \sum_{\alpha} \frac{m_{\alpha}}{2} \dot{r}^2_{\alpha}\dot{\vek{r}}_{\alpha}
\delta(\vek{r}-\vek{r}_{\alpha}(t)).
\end{align}
The force $m\ddot{\vek{r}}$ appearing in the time derivative of $U_m$
is just the Lorentz force $\vek{F}=e(\vek{E}+ \dot{\vek{r}} \times
\vek{B})$, yielding
\begin{align}
\begin{split}
\label{pt-mechanical}
\frac{\partial}{\partial t} U_m(\vek{r},t) + \div \vek{S}_m (\vek{r},t) &=
\vek{j}(\vek{r},t)\vek{E}(\vek{r},t).
\end{split}
\end{align}
This can finally be combined with eq.~\eqref{pt-em} to Poynting's theorem as
shown in eq.~\eqref{pt-general}. No restrictions have been made to the behaviour of
the fields or particles or the nature of the latter, except for non-relativistic
velocities. This approach will turn out to be inevitable for establishing a
relation between electromagnetic and mechanical energy. Mechanical energy here
is purely kinetic, any potential energy is attributed to the fields.

The continuity of the energy flux at medium boundaries is an aspect to which
much attention is paid in literature. 
The electromagnetic Poynting vector $\vek S_e$ derived here is
not necessarily continuous, but this does not account for physical
meaningfulness. It is the energy continuity eq.~\eqref{pt-general} that holds. The
continuity of one the quantities involved depends on that of the
others, which in turn is subject to the system and model considered.

\section{Material equations and susceptibility}
The macroscopic Maxwell equations originally were introduced to obtain a
global, effective description of electrodynamics in the presence of a medium.
One may consider their form a well-founded historic convention, which can be
reproduced simply by splitting $\vek j = \jind + \jext$ and the ansatz
\begin{align}
	\label{jind-rotm}
	\jind = \dot{\P} + \rot \M,
\end{align}
which is also suggested \cite{Nolting} by the well-known multipole expansion of
the electromagnetic properties of the medium \cite{Jackson}, 
and the respective \emph{material equations}
\begin{subequations}
\label{material-eq}
\begin{align}
    \D(\vek r, t) &= \varepsilon_0 \E(\vek r, t) + \P(\vek r, t)\\
    \H(\vek r, t) &= \frac{1}{\mu_0} \B(\vek r, t) - \M(\vek r, t)\\
	\label{eq-p-sus}
    \P(\vek r, t) &= \varepsilon_0 \int \upd^3 \vek{r}' \upd t'\;
    \tens{\chi}\elec (\vek r, \vek r', t, t') \E(\vek r', t')\\
	\label{eq-m-sus}
    \M(\vek r, t) &= \frac{1}{\mu_0} \int \upd^3 \vek{r}' \upd t' \;
    \tens{\chi}\magn (\vek r, \vek r', t, t') \B(\vek r', t').
\end{align}
\end{subequations}

Note the clear analogy in this notation and the obvious distinction of
fields $\E,\B$ and derived fields $\D,\H$. Usually, textbooks
\cite{Landau,Born2003,Nolting} focus more on the analogy
\begin{align}
\label{bh-hist-symmetry}
\B=\mu_0\tens\mu_r\H, \quad \D=\varepsilon_0\tens\varepsilon_r\E,
\end{align} 
\ie, with constant tensorial relations \eqref{eq-p-sus}, \eqref{eq-m-sus},
and consequently treat $\E$ and $\H$ on the same level. But the physically
relevant fields are actually $\E$ and $\B$, because they determine, \eg, the
Lorentz force, the energy density, and the electromagnetic energy flux. 

Neither the multipole expansion discussion in \cite{Jackson} nor the tensorial
descriptions in \cite{Landau,Born2003} end up with non-local or inhomogeneous
material equations and thus cannot be used with spatial dispersion and bounded
media. 
Hence, a more general dependency of the derived fields on $\E$ and $\B$, as
shown in eqs.~\eqref{material-eq}, needs to be used. Here, the tensorial
susceptibilities $\tens\chi\elec$, $\tens\chi\magn$ represent exactly the
linear properties of an arbitrarily complex stationary medium. However, 
PT can still be derived in the same manner without any restrictions to the
electromagnetic fields, properties of the 
medium or mathematical properties of $\tens\chi\elec$ or $\tens\chi\magn$,
except for the temporal homogeneity of the latter. 
Also, eq.~\eqref{eq-vekpoynting-e-old} can still be obtained, but then $\vek j$
does no longer fulfil eq.~\eqref{particle-current} and thus cannot be coupled with
the mechanical energy continuity relation \eqref{pt-mechanical}, and the field
energy density cannot be given.

In optics, the approximation $\tens\chi\magn \equiv 0$ is widely
accepted. Then, $\vek{S}_{e}$ becomes $\E\times\H$ and $\jind =
\dot\P$. From a higher level of abstraction one can also \emph{postulate} 
that $\jind = \dot\P$. This avoids pinpointing magnetization and other 
difficulties and results in the following material equations \cite{Halewi1992}:
\begin{subequations}
\label{material-eq-opt}
\begin{align}
    \D(\vek r, t) &= \varepsilon_0 \E(\vek r, t) + \P(\vek r, t)\\
    \P(\vek r, t) &= \varepsilon_0 \int \upd^3 \vek{r}' \upd t'\;
    \tens{\chi} (\vek r, \vek r', t, t') \E(\vek r', t')\\
    \H(\vek r, t) &= \frac{1}{\mu_0} \B(\vek r, t)
\end{align}
\end{subequations}
Now, the $\H$ field is merely a rescaled $\B$ field and any
electromagnetic effects are contained in $\P$.

Much of the controversy around the validity of Poynting's theorem
may be due to inconsistent formulation of the material equations or using
eq.~\eqref{eq-vekpoynting-e-old} despite of magnetization. 
In \cite{NelsonGeneralizingPoynting}, the author correctly finds that
eq.~\eqref{eq-vekpoynting-e-old} together with the common approximation
\eqref{bh-hist-symmetry} for the material equations does not apply 
for wide categories of interactions, such as spatial dispersion. He also
bases his further considerations on the Maxwell-Lorentz equations 
and equally finds eq.~\eqref{pt-em} but then develops an elaborate expression for
$\vek j \E$ within the (local) multipole expansion approximation. 
The result, of course, is
an expression for our $\vek S_m$ in eq.~\eqref{pt-general} (compare 
\cite[eq.~21]{NelsonGeneralizingPoynting}). This is an honorable effort, 
even though his results can only hold true in the limits of this approximation.
Unfortunately, the author does not identify $(\E\times\B)/\mu_0$ as a valid
electromagnetic Poynting vector and eq.~\eqref{pt-general} as a more general
energy conservation law.

\section{The work of Bishop/Maradudin and its conceptual problems}
In \cite{BishopMaradudin76}, Bishop and Maradudin analyze the energy flow
in a semi-infinite spatially dispersive dielectric (\ie, \emph{half-space
geometry}) on which light is incident from the vacuum.
The energy flux in the vacuum is given by the electromagnetic Poynting vector
$\vek{S}_e$ only. In the medium, 
the mechanical Poynting vector $\vek{S}_m$ appears additionally. They state that
the DA fails to conserve energy by showing that
$\vek{S}_m$ does not evolve continuously from zero at the surface, as if there
was an energy source.
For this purpose, they develop a model of the transport of
mechanical energy in a special system, a diatomic cubic crystal, on the
basis of fixed, undamped and uncoupled harmonic oscillators only, then derive
PT from Maxwell's equations and adapt it for their model. 
In \cite[eq.~4.7]{BishopMaradudin76} they postulate that $\vek{S}_m$ be represented by the
oscillations; the spatial restriction to the medium is forced by an appropriate
step function. 

This postulate does not necessarily fulfill the continuity
condition. $\vek{S}_m$ is continuous only if the oscillator
amplitude vanishes at the surface. One of the possibilities to ensure this is 
Pekar's ABC \cite{pekar-abc}, as the authors confirm in their conclusion.
It is no surprise that the step function leads to a delta-shaped 
term \cite[eq.~4.10]{BishopMaradudin76} 
when conducting the divergence operation in eq.~\eqref{pt-general}. 
Of course, such a source term is not compatible with energy conservation.
Consequently, the authors generalize their model and derive additional
conditions for its free parameters.

This kind of approach is highly doubtful because any predictions 
might have to be attributed to the model and its deficiencies. Here, this is
clearly the case: The mechanical Poynting vector as postulated does neither
contain all mechanical energy transport in the medium 
(\eg, via phonons, center-of-mass motion of excitons etc.) nor does it fulfill
itself the condition to be analyzed.

\section{Compatibility of DA and PT}
We rather suggest an approach free of models. It includes but is not limited to
the case of DA; it does not make any assumptions on the microscopic nature of
the medium but instead is based on the susceptibility $\chi$ as an exact but unknown
representation of the latter.

We regard a stationary isotropic medium in the \emph{slab geometry}. Light is
incident perpendicular to the medium surface. Then Poynting's theorem reduces
to:
\begin{align}
\begin{split}
\label{pt-slab-stationary}
\frac{\partial S_e(x,t)}{\partial x} = -\frac{\partial S_m(x,t)}{\partial x} = -j(x,t)E(x,t) =: - W(x,t)
\end{split}
\end{align}
By expressing $E$ and $B$ through the vector potential 
$A(x,t)$, $j$ through $\partial P/\partial t$ and Fourier
transforming $t \to \omega$, we obtain
\begin{align}
\begin{split}
\label{dissipation-stationaer}
W(x,\omega) &= \ii \varepsilon_0 \int \frac{\upd\omega'}{2\pi}
\omega'^2(\omega-\omega') \\
& \cdot \int \upd x' \chi(x,x',\omega')A(x',\omega')A(x,\omega-\omega').
\end{split}
\end{align}
There is no need for taking a time average.
As $\chi=0$ in vaccuum, it may be
replaced by $\chi(x,x',\omega) = \Theta(\Lh - |x|)\Theta(\Lh -
|x'|)\chi(x,x',\omega)$, where $L$ is the length of the slab. Of course, this
is also valid for the special case of DA, where $\chi$ just simplifies to
$\Theta(\Lh - |x|)\Theta(\Lh - |x'|)\chi(x-x',\omega)$. Evaluating the step
functions, we have:
\begin{align}
\nonumber
\begin{split}
W(x,\omega) &=  \ii \varepsilon_0 \int \frac{\upd\omega'}{2\pi}
\omega'^2(\omega-\omega')\Theta(\Lh - |x|)A(x,\omega-\omega') \\
& \quad \quad \cdot
\int_{-\Lh}^{\Lh} \upd x' \, \chi(x,x',\omega')A(x',\omega')\\
&=: \Theta(\Lh - |x|) \bar{W}(x,\omega)
\end{split}
\end{align}
On the left surface, the step function reduces to $\Theta(\Lh + x)$. 
We now obtain a new, model-free relation for
$S_m$ by integrating \eqref{pt-slab-stationary}:
\begin{align}
\begin{split}
\label{sm-by-w}
S_m(x,\omega) & = \Theta(\Lh + x)\int_{-\Lh}^{x} \upd x' \, \bar{W}(x',\omega)
\end{split}
\end{align}
As can clearly be seen, $S_m$ is zero at the surface and grows continuously for
$x>-\Lh$, because $\bar{W}$ is a finite value. It is therefore continuous at the
surface. There is no evidence for any energy source or violation of Poynting's
theorem.

\section{Energy conservation with bulk susceptibility and ABC's}

Notwithstanding the first successful microscopic non-local computation of
the polarisation \cite{Jahnke2000}, which required supercomputers, 
the pragmatic assumption of a bulk susceptibility $\chi(q,\omega)$ is still
widely used to analyze light propagation through bounded media. 
In this case, the dispersion relation 
\begin{align}
\begin{split}
q(\omega)^2 = \frac{\omega^2}{c^2}(1 + \chi(q,\omega))
\end{split}
\end{align}
can have various solutions $q_i(\omega)$, giving rise for different
\emph{polariton modes}. Consequently, the electromagnetic
field inside is a superposition of these forward and backward propagating
modes (we stick with slab geometry):
\begin{align}
\begin{split}
E(x,\omega) = \sum_i \left[ f_i(\omega)\e^{\ii q_i(\omega) x} +
b_i(\omega)\e^{-\ii q_i(\omega) x} \right]  
\end{split}
\end{align}

The \emph{extinction theorem}, 
\begin{align}
\label{absorption-matrix}
a_1(\omega) &= 1 - \left|r\right|^2  - \left|t\right|^2,
\end{align}
where $a_1$ is the absorption and $r,t$ the coefficients of reflected or
transmitted light, resp., 
and the MBC's,
\begin{subequations}
\begin{align}
\label{mbc-1}
\e^{-\ii q_0 \Lh} + r\e^{\ii q_0 \Lh} &= \sum_i \left( f_i \e^{-\ii q_i\Lh} +
b_i \e^{\ii q_i \Lh} \right), \\
\label{mbc-2}
\e^{-\ii q_0 \Lh} - r\e^{\ii q_0 \Lh} &= \sum_i \frac{q_i}{q_0}\left( f_i
\e^{-\ii q_i\Lh} - b_i \e^{\ii q_i \Lh} \right), \\
\label{mbc-3}
t\e^{\ii q_0 \Lh} &= \sum_i \left( f_i \e^{\ii q_i\Lh} + b_i \e^{-\ii q_i\Lh}
\right), \\
\label{mbc-4}
t\e^{\ii q_0 \Lh} &= \sum_i \frac{q_i}{q_0}\left( f_i \e^{\ii q_i \Lh} + b_i
\e^{-\ii q_i \Lh} \right),
\end{align}
\end{subequations}
where $q_0=\omega/c$, 
form a set which could be solved if only a single mode was present but 
in fact is under-determined due to the polariton modes. The postulation of 
ABC's help solving this problem, but some of them were known to conflict with
energy conservation. (For a concise summary on ABC's and the controversy around
them, see \cite{Venger2004} and refs.\ therein). Most of these ABC's, as well as the
DA, can be reproduced by \emph{generalized ABC's} \cite{Halewi1992}.

Using the RHS of the integral representation of eq.~\eqref{pt-slab-stationary}, an
independent expression for the absorption can be derived:
\begin{align}
\label{absorption-henneb}
a_2(\omega) = \frac{1}{2 \ii}\frac{\omega}{c} \int \upd x \upd x' \,
A^*(x,\omega)\hat\chi(x,x',\omega)A(x',\omega), \\
\text{with}\quad \hat\chi(x,x',\omega) = \chi(x, x',\omega) - \chi^*(x',x,\omega)
\end{align}
It follows by applying a monochromatic wave and some straightforward
calculation. Then, static terms ($\propto \delta(\omega)$) and such $\propto
\delta(\omega \pm 2\omega_0)$ appear. The former can be regarded seperately,
elegantly circumventing cycle averaging. 
The important relations \eqref{absorption-matrix} and \eqref{absorption-henneb}
are presented in \cite{planckslaw} in a quantum mechanical generalization
valid for non-locally dispersive media and oblique incidence. They take almost
the same form as shown here.

The polarisation here follows from eqs.~\eqref{material-eq-opt} as
\begin{equation}
\nonumber
\begin{split}
P(x,\omega)=
\sum_{j}\chi_j(\omega)\left[f_j(\omega)\e^{\ii q_j(\omega)x} +
b_j(\omega)\e^{-\ii q_j(\omega)x} \right].
\end{split}
\end{equation}
It can be used to calculate $a_2(\omega)$ in 
a variant of eq.~\eqref{absorption-henneb} based on field strength and 
polarisation function only,
\begin{align}
a_2(\omega) = \frac{\omega}{c} \Im \int_{-\Lh}^{\Lh} \upd x \, 
E^*(x,\omega)P(x,\omega).
\end{align}
Now, constructing eq.~\eqref{absorption-matrix} as
\begin{align}
\begin{split}
a_1(\omega) = \frac{1}{4}\big(
&\left|\eqref{mbc-1}+\eqref{mbc-2} \right|^2 -
\left|\eqref{mbc-1}-\eqref{mbc-2} \right|^2 
\\
-&
\left|\eqref{mbc-3}+\eqref{mbc-4} \right|^2 +
\left|\eqref{mbc-3}-\eqref{mbc-4} \right|^2 
\big),
\end{split}
\end{align}
the last term equaling zero,
it can be shown that $a_1=a_2$. As no ABC's have been used up to now, this means
that energy conservation in such a system is ensured by the MBC's
alone.

\section{Conclusion}
We hopefully could present the concept of Poynting's theorem in
conjunction with a mechanical energy flux vector in an illustrative way,
concretizing and extending textbooks in this sense, offer a best-practice
approach, and show that doubts on its validity in certain conditions, such as
spatial dispersion, are unsubstantiated. By the given approach, a clear relation 
between electromagnetic and kinetic energy in a system can be established, 
and electromagnetic energy density and flux are determined 
by the fundamental physical fields $\E$ and $\B$ alone.

We then have proven analytically that energy conservation in the propagation of
light through bounded media is fulfilled for arbitrary susceptibilities, and
thus also for the dielectric approximation. 

Often, the use of models and approximations for the microscopic nature of a
medium is misleading, and any predictions have to be taken with
a grain of salt. This was pointed out as a problem in \cite{BishopMaradudin76}
and in the common perception of Poynting's theorem in general. 
Great care must be taken in the application of the material
equations, and it seems advisable to always start from \eqref{material-eq} and 
the Maxwell-Lorentz equations \eqref{maxwell-lorentz}, especially when spatial
dispersion is dealt with. If a model for the mechanical Poynting vector
is applied, it must be consistent with the susceptibility. Namely, it can be
constructed via eq.~\eqref{sm-by-w}.

Furthermore, we were able to prove analytically that energy conservation in
approximations employing bulk susceptibility is completely assured by Maxwell's boundary 
conditions alone, even if multiple polariton modes are predicted. Contrary to what
was thought before, additional boundary conditions do not play any role in this
respect, whether positive or negative. 
Consequently, from the energy conservation
point of view, it is possible to freely choose the ABC's that match microscopic
calculation results best (which would probably be a combination of Pekar's and a
\emph{dead layer}, see \cite{MuljarovZimmermann}).

All these results save a plethora of recent and older work the shortcoming
of possibly violating fundamental principles, and the use of their results 
in research is finally justifiable, at least in this respect.

\begin{acknowledgments}
The authors would like to thank the \emph{Deutsche Forschungs\-gemein\-schaft} for
support through \emph{Sonder\-forschungs\-bereich 652}.
\end{acknowledgments}

\end{document}